# Giant magnetic response of a two-dimensional antiferromagnet

Lin Hao[1], D. Meyers[2], Hidemaro Suwa[1], Junyi Yang[1], Clayton Frederick[1], Tamene R. Dasa[3], Gilberto Fabbris[4], Lukas Horak[5], Dominik Kriegner[5,6], Yongseong Choi[4], Jong-Woo Kim[4], Daniel Haskel[4], Philip J. Ryan[4,7], Haixuan Xu[3,*], Cristian D. Batista[1,8,*], M. P. M. Dean[2,*], Jian Liu[1,*]

[1] Department of Physics and Astronomy, University of Tennessee, Knoxville, Tennessee 37996, USA

[2] Department of Condensed Matter Physics and Materials Science, Brookhaven National Laboratory, Upton, New York 11973, USA

[3] Department of Materials Science, University of Tennessee, Knoxville, Tennessee 37996, USA

[4] Advanced Photon Source, Argonne National Laboratory, Argonne, Illinois 60439, USA

[5] Department of Condensed Matter Physics, Charles University, Ke Karlovu 3, Prague 12116, Czech Republic

[6] Institute of Physics, Academy of Sciences of the Czech Republic, v.v.i., Cukrovarnická 10, 16253 Praha 6, Czech Republic

[7] School of Physical Sciences, Dublin City University, Dublin 9, Ireland

[8] Quantum Condensed Matter Division and Shull-Wollan Center, Oak Ridge National Laboratory, Oak Ridge, Tennessee 37831, USA

Correspondence to: H.X. (hxu8@utk.edu), C.D.B. (cbatist2@utk.edu), M.P.M.D. (mdean@bnl.gov), J.L. (jianliu@utk.edu).

**A fundamental difference between antiferromagnets and ferromagnets is the lack of linear coupling to a uniform magnetic field due to the staggered order parameter[1]. Such coupling is possible via the Dzyaloshinskii-Moriya (DM) interaction[2,3] but at the expense of reduced antiferromagnetic (AFM) susceptibility due to the canting-induced spin anisotropy[4]. We solve this long-standing problem with a top-down approach that utilizes spin-orbit coupling in the presence of a hidden SU(2) symmetry. We demonstrate giant AFM responses to sub-Tesla external fields by exploiting the extremely strong two-dimensional critical fluctuations preserved under a symmetry-invariant exchange anisotropy, which is built into a square-lattice artificially synthesized as a superlattice of $SrIrO_3$ and $SrTiO_3$. The observed field-induced logarithmic increase of the ordering temperature enables highly efficient control of the AFM order. As antiferromagnets promise to afford switching speed and storage security far beyond ferromagnets[5-8], our symmetry-invariant approach unleashes the great potential of functional antiferromagnets.**

Low-dimensional antiferromagnets, exemplified by high-$T_c$ cuprates, are known for extremely rich emergent behaviors, such as unconventional superconductivity, exotic magnetism, magnon condensates, quantum phase transitions and criticality[9-11]. According to the Mermin-Wagner theorem[12], strong critical fluctuations of an isotropic 2D antiferromagnet prohibit long-range magnetic ordering at finite temperatures, and lead to a giant AFM susceptibilty $\chi_{AF}$ due to the exponentially diverging magnetic correlation length $\xi \propto e^{2\pi\rho_s/T}$ (where $\rho_s$ is the stiffness) as $T \to 0$[13]. While it is well known that magnetic field can suppress the fluctuations, the induced Zeeman energy must be comparable to the AFM interaction to significantly enhance the ordering stability[14]. As a result, the field required for a siziable effect is often very large and even unpractical. The underlying limitation originates from the fact that the AFM order is a locked pair of opposite interpenetrating ferromagnetic sublattices[15] (Fig. 1a) and only couples weakly to magnetic field via a quadratic term.

Although symmetrically forbidden in a collinear state, the linear coupling of the AFM order parameter (OP)[5] to a uniform magnetic field can be enabled in the presence of spin canting, which is caused by the antisymmetric anisotropic exchange - the well-known DM interaction[2,3] $\vec{D}_{ij} \cdot \vec{S}_i \times \vec{S}_j$ between neighboring spins (Fig. 1b). The canting creates a small net moment and allows the external field to linearly drive the AFM order as an effective staggered field. The DM interaction is, however, also accompanied by a symmetric anisotropic exchange $\vec{S}_i \cdot \overleftrightarrow{\delta} \cdot \vec{S}_j$ term[2]. Both antisymmetric and symmetric anisotropic exchanges necessarily induce local spin anisotropy[16] (Fig. 1b). This route thus presents a dilemma in that the enforced magnetic axis confines the AFM spins, which would otherwise be free to rotate (Fig. 1a), and reduces the AFM susceptibility $\chi_{AF}$.

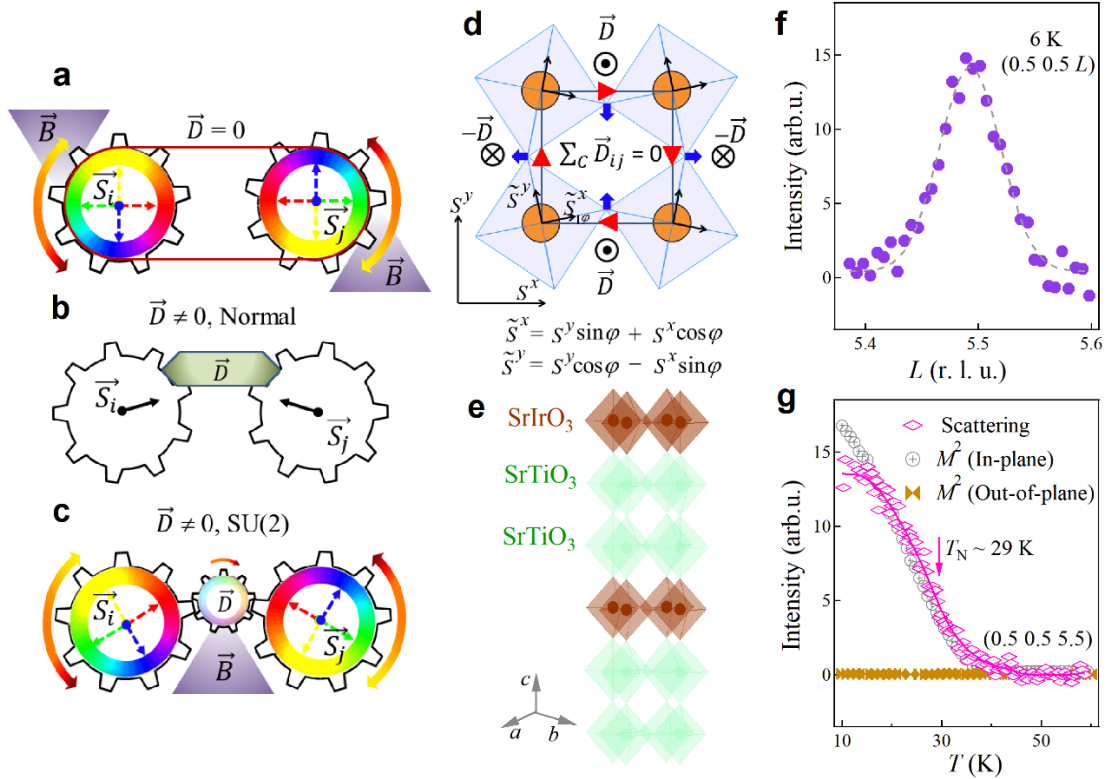

**Figure 1 | Design and realization of spin canting without spin anisotropy via a SU(2)-invariant DM interaction. a-c,** Schematic diagrams of a pair of antiferromagnetically coupled spins. The pair is fully antiparallel and free to rotate together in all directions (**a**). Under a typical DM interaction (**b**), the pair is canted toward a preferential orientation, which is stable against magnetic field. If the DM interaction preserves the rotational symmetry (**c**), the pair is again highly susceptible to magnetic field via the canting. **d,** A square lattice where DM interactions caused by planar octahedral rotation (blue arrows) preserves SU(2) symmetry. The spins $\vec{S}$ can be mapped on a local spin frame $\vec{\tilde{S}}$ according to the shown transformation. Red arrows denote the summation loop of DM vectors. **e,** Layered structure of the superlattice. **f,** Reciprocal space *L*-scan across the (0.5 0.5 5.5) magnetic reflection at the Ir $L_3$-edge and 6 K. **g,** Temperature-dependence of the AFM Bragg peak intensity at zero field reveals $T_N \sim 29$ K, defined as the maximum slope of the AFM OP. A similar onset behavior is seen in the in-plane remanent magnetization, which is plotted squared since scattering is proportional to the OP squared[26]. The out-of-plane component was also shown for comparison.

While enabling a strong linear coupling and preserving the 2D AFM susceptibility seem fundamentally incompatible, we show here that a solution is possible if the spin isotropy is protected by the global symmetry of the system under the local anisotropic exchanges (Fig. 1c). Such a symmetry-invariant exchange anisotropy was first proposed more than two decades ago in the context of superexchange pathways in spin-half AFM square lattices[16], but, to the best of our knowledge, has not been experimentally realized or utilized. The effective Hamiltonian led by an AFM Heisenberg interaction $J$ between the neighboring moments is

$$H = \sum_{<i,j>}[J\vec{S}_i \cdot \vec{S}_j + \vec{D}_{ij} \cdot \vec{S}_i \times \vec{S}_j + \delta S_i^z S_j^z] - h \sum_i S_i^x \quad (1),$$

where $<i, j>$ runs over all neighboring pairs. The last term of equation (1) represents the Zeeman energy in a uniform magnetic field $B$ along the $x$-axis with $h = g_{aa}\mu_B B$, where $g_{aa} \approx -2$ is the g-factor, and $\mu_B$ is the Bohr magneton. This term vanishes for a collinear antiferromagnet, but is finite for a canted AFM order with nonzero $\vec{D}_{ij}$, when the local inversion symmetry is broken by an in-plane octahedral rotation of the 2D corner-sharing octahedral network (Fig. 1d) to enable both the DM interaction and symmetric anisotropic exchange. While such exchange anisotropy usually suppresses the large AFM susceptibility from an isotropic 2D Heisenberg model, it was shown that[16], if the octahedral rotation is purely in-plane (Fig. 1d), the exchange anisotropy will preserve the continuous SU(2) spin symmetry. Specifically, this condition guarantees that $\delta = \sqrt{J^2 + D^2} - J$ with $D = |\vec{D}_{ij}|$ and the DM vectors $\vec{D}_{ij}$ are all perpendicular to the basal plane and alternate their signs when circulating around a closed loop (Fig. 1d). As a result, equation (1) acquires a hidden SU(2) symmetry at zero-field. The hidden SU(2) is unveiled by a *staggered z-axis* rotation of the local spin reference frame by the canting angle $\varphi$ ($\tan 2\varphi = D/J$)[16] (Fig. 1d). In the new frame, equation (1) recovers an isotropic 2D Heisenberg model (Supplementary Information 5)

$$H = \sum_{<i,j>} \tilde{J}\vec{\tilde{S}}_i \cdot \vec{\tilde{S}}_j - h \cos\varphi \sum_i \tilde{S}_i^x + h \sin\varphi \sum_i e^{i\mathbf{Q}\cdot\mathbf{r}_i}\tilde{S}_i^y \quad (2).$$

Here $\tilde{J} = \sqrt{J^2 + D^2}$, and $\mathbf{Q} = (\pi, \pi)$ is the AFM ordering wave vector. Key to this hidden continuous symmetry is the so-called "unfrustrated condition" $\sum_C \vec{D}_{ij} = 0$ [16] for the DM vectors (Fig. 1d), where $C$ denotes any closed loop on the square lattice. Fundamentally, this condition is achieved here because the global structural point group symmetry of the square lattice is preserved in the presence of the in-plane octahedral rotations. It is noteworthy that the hidden SU(2) symmetry holds not only in the effective spin model, but also the high-energy single-band Hubbard model including the charge degrees of freedom[16]. The SU(2)-invariant DM interaction is remarkable in that the large $\chi_{AF}$ is not only preserved but also manifests under $h \ll J$ due to the large linear coupling with the AFM OP $\widetilde{M}_{st} = \sum_i e^{i\mathbf{Q}\cdot\mathbf{r}_i} \langle \widetilde{S}_i^y \rangle$ unveiled in equation (2).

To exploit the response of the hidden SU(2) symmetry to the external field, we employ the epitaxial superlattices of $(SrIrO_3)_1/(SrTiO_3)_2$ grown along the pseudocubic [001]-direction on $SrTiO_3$ substrates[17] (Fig. 1e). The design utilizes the magnetic degrees of freedom arising from Kramers doublets of the $Ir^{4+}$ $5d^5$ ions[18]. As found in a variety of iridate compounds[9,19,20], these doublets result from the splitting of the active $t_{2g}$ levels caused by a large spin-orbit coupling ~0.4 eV and they can be represented with effective $S = 1/2$ pseudospins (Supplementary Information 4). In this situation, even a weak Coulomb repulsion is sufficient to generate a 2D AFM Mott insulating state, such as that in $Sr_2IrO_4$[19]. On the other hand, unlike ordinary $S = 1/2$ spins present in lighter transition metal oxides, the effective $S = 1/2$ pseudospins have far stronger spin-orbit coupling which leads to much larger DM interactions and spin canting commonly found in magnetic iridates[18,21,22]. Indeed, the ground state of the confined $IrO_6$ octahedral layer in our superlattice is revealed as a 2D antiferromagnet with a Néel transition at $T_N$ ~29 K and significant canted moments (Figs. 1f and g), which is essential for the hidden SU(2) symmetry. The reason for using a bilayer $SrTiO_3$ spacer here is two-fold. Firstly, to realize the unfrustrated condition, we performed density functional theory calculation and found only the in-plane octahedral rotation exists, whereas out-of-plane rotation would occur when the spacer is thinner (Supplementary Information 1). We confirm this $D_4$-symmetric structure by synchrotron x-ray diffraction (Supplementary Information 2), fulfilling the unfrustrated condition. Such an octahedral rotation pattern is also consistent with the observed in-plane canted moment, similar to the $Sr_2IrO_4$ with a canting angle $\varphi \sim 10°$[18,23], and the absence of out-of-plane net magnetization (Fig. 1g). Other layered perovskites such as $La_2CuO_4$ tend to have out-of-plane octahedral roations[16]. Additionally, increasing the interlayer spacing reduces the interlayer exchange $J_\perp$. This is crucial because $J_\perp$ stabilizes $T_N$ at the cost of reducing the $\chi_{AF}$[24] and it must be orders of magnitude smaller than $h$ in order to exploit the symmetry-invariant DM interaction of a quasi-2D system. Indeed, when decreasing the $SrTiO_3$ spacer from a bi-layer to a single-layer, $T_N$ is increased from ~29 K (Fig. 1g) to ~140 K[17,25], consistent with a strong reduction of $J_\perp$.

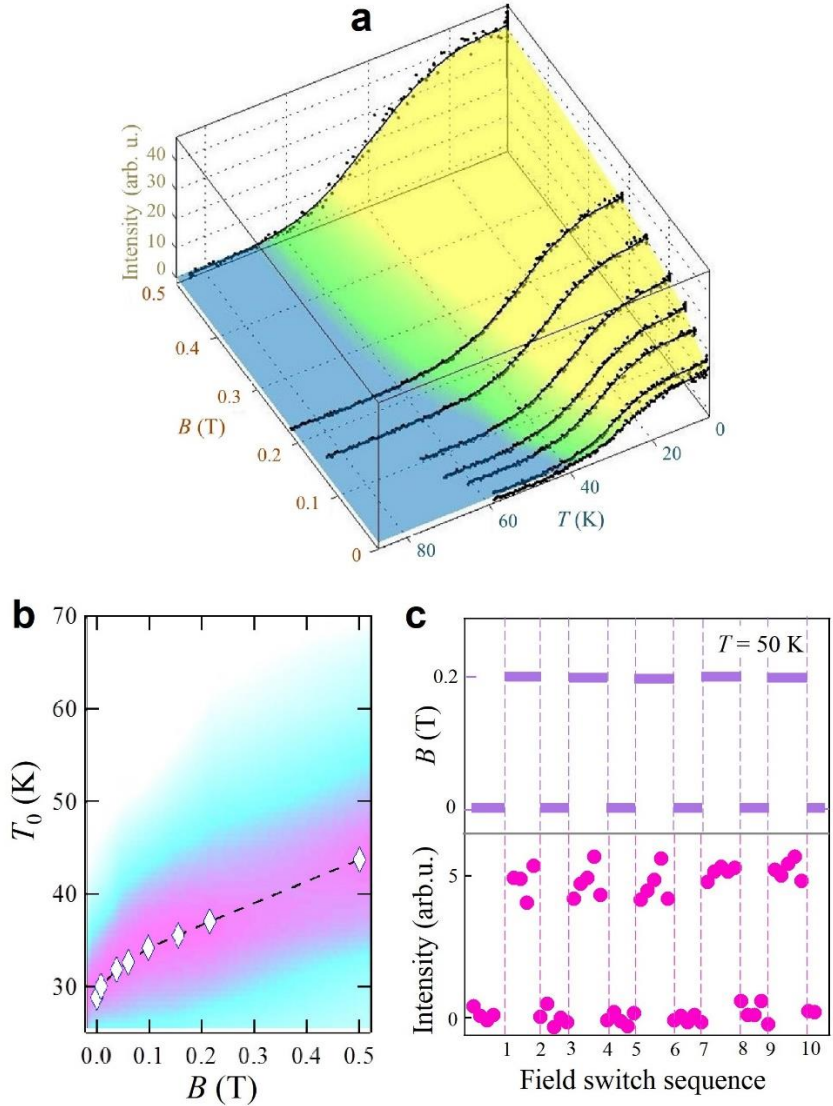

**Figure 2** | **Magnetic diffraction in applied magnetic fields. a,** Temperature dependence of the (0.5 0.5 5.5) AFM Bragg peak under various in-plane magnetic fields. The colored surface mesh highlights the dramatic increase of the temperature boundary below which the magnetic peak becomes observable. **b,** Relation between crossover temperature $T_0$ and magnetic field $B$. The color scale highlights the crossover region. **c,** The magnetic peak intensity in response to an on/off 0.2 T field-switching sequence at 50 K.

Having verified the realization of the hidden SU(2) symmetry, we explored the response of the AFM transition to an in-plane magnetic field. As can be seen from Fig. 2, the thermal stability of the AFM order is rapidly enhanced as the field increases from 0 to 0.5 T. For instance, while the AFM Bragg peak onsets at ~40 K at zero field, it is readily observable below 70 K at just 0.5 T. Since the zero-field Néel transition becomes a crossover under external field due to the linear coupling, the crossover temperature, $T_0$, is defined similarly to $T_N$ as the temperature that maximizes the slope of the OP extracted from the peak intensity[26]. Figure 2b shows the drastic enhancement of $T_0$, especially at small fields near 0.1 T, displaying a logarithmic behavior (Fig. 3a). The enhancement of $T_0$ at 0.5 T is ~50%, which is remarkable considering that the Zeeman energy at this maximum applied field is still three orders of magnitude smaller than $J$~50 meV[20,27,28] (Fig. 3a). The extreme sensitivity of $T_0$ enables complete on/off switching of the AFM order with small magnetic fields. Figure 2c shows the *in-situ* observation of the AFM Bragg peak at 50 K. Zero counts are observed when the field is off because there is no AFM order at this temperature. In contrast, clear magnetic peak intensity is detected when the field is on, indicating activation of the AFM long-range order. The switching is highly reliable as evidenced from the reproducible and prompt response of the peak intensity even after turning the magnetic field on/off multiple times.

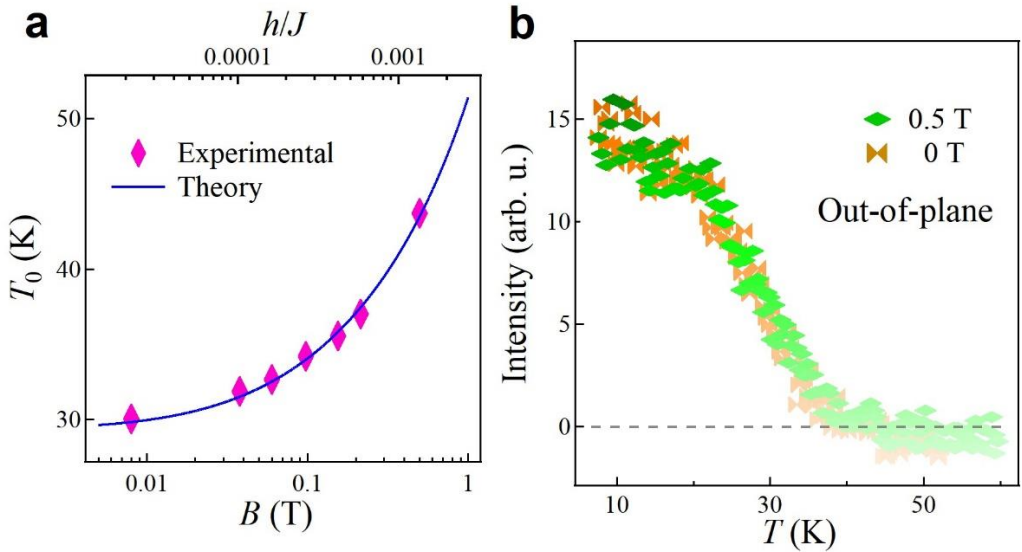

**Figure 3 | Theoretical analysis and experimental confirmation. a,** Comparison between the measured field-dependence of the crossover temperature $T_0$ and the logarithmic increase [equation (3)] expected for a quasi-2D version of equation (1). $J$ is set as 50 meV[20,27,28] to normalize $h$ (Methods). **b,** Temperature scan of the (0.5 0.5 5.5) magnetic peak under a 0.5 T out-of-plane ($B//z$) field is shown together with a zero-field scan done under the same conditions.

The rapid increase of $T_0(B)$ can be quantitatively accounted for by equation (2) with addition of the small higher-order exchange anisotropies, $H' = -\Gamma_1 \tilde{S}_i^z \tilde{S}_j^z$, induced by the small Hund's coupling of the Ir$^{4+}$ ion[18,20]. This term is responsible for easy-plane anisotropy and is $\sim 10^{-4} J$ (Methods). Although the Hamiltonian $(H + H')$ lowers the SU(2) symmetry to U(1), the resulting planar continuous symmetry still leads to an exponentially divergent antiferromagnetic correlation length in the vicinity of a Berezinskii–Kosterlitz–Thouless (BKT) transition[29]: $\xi \sim e^{b/\sqrt{t}}$, where $t = \frac{T - T_{BKT}}{T_{BKT}}$ and $b$ is a $\Gamma_1$ dependent constant. This essential singularity of the transition point renders the AFM order highly susceptible to the external field. We then treat $h$ and $J_\perp$ as perturbations to $(H + H')$. These perturbations are negligible in the $T \gg T_0$ regime dominated by in-plane vortex-antivortex excitations. The crossover to the regime characterized by a large AFM OP occurs at the temperature scale $T_0$, where the combined cost from $h$ and $J_\perp$ for separating an in-plane vortex-antivortex pair by a distance $\xi$ is comparable to the intralayer exchange energy[29]: $(2|J_\perp|S^2 + S|h \sin \varphi|)\xi^2 \sim \tilde{J}S^2 \ln \xi$. The resulting crossover temperature is

$$T_0 = T_{BKT} + \frac{4b^2 T_{BKT}}{\left[\ln\left(\frac{c\tilde{J}S^2}{2|J_\perp|S^2 + S|h \sin \varphi|}\right)\right]^2} \quad (3),$$

where $c$ is a constant accounting for the effects of quantum fluctuations and disorder. At zero field, the finite $J_\perp$ turns the BKT transition to a Neel transition, i.e., $T_N = T_0(h=0)$. The rapid increase of $T_0$ arises from the logarithmic dependence on the magnetic field, which linearly couples to the AFM order. To fit the observed $T_0(B)$ with equation (3), we estimated the constant $b$ from the crossover temperature calculated with classical Monte Carlo simulations (Methods). Figure 3a shows that the experimental data is well explained by equation (3). The estimated $J_\perp \sim 10^{-3}$ meV is two orders smaller than the Zeeman energy at 0.5 T.

In the absence of DM interaction, the linear field-OP coupling in equation (3) would have to be replaced by the much weaker quadratic coupling that gives an energy contribution two orders of magnitude smaller than the interlayer interaction for a 0.5 T field, i.e., to a negligibly small field-induced increase of $T_0$ (Methods). We confirmed this picture by applying a 0.5 T *out-of-plane* field ($B//z$) that has no coupling to the *in-plane* spin canting. The measured temperature-dependence of the magnetic Bragg peak indeed shows no observable change of $T_0$ compared with $T_N$ (Fig. 3b). As a comparison, similarly small effects have been seen in Cu$^{2+}$-based quasi-2D materials with canted moment that is two orders smaller than iridates, and a significant increase of $T_0/T_N$ therein demands a much larger field to match the AFM exchange[14].

Our study shows that a SU(2)-invariant DM interaction can enable an unprecedented control of AFM order by a small magnetic field. This mechanism drives a logarithmic increase of the ordering temperature of a 2D antiferromagnet by exploiting the large 2D critical fluctuations under a hidden continuous symmetry. Engaging with the hidden SU(2) symmetry may lead to dramatic effects, pointing to rich spin-orbit physics of iridates beyond the high-$T_c$ analogy. Since symmetry-invariant exchange anisotropy is not restricted to square lattice, the demonstration of this concept is expected to facilitate development of new antiferroic systems and devices with improved efficiency.

## METHODS

**First-principles density functional calculations.** Density Functional Theory (DFT) calculations were performed using projector-augmented wave method with the generalized gradient approximation (GGA)[30] as implemented in the Vienna *ab-initio* Simulation Package[31]. The plane wave cutoff was chosen as 500 eV based on the convergence tests. To model the epitaxial relationship of samples grown on a SrTiO$_3$ substrate, the in-plane lattice parameters were fixed to the substrate value, 3.905 Å. A 4 × 4 × 3 Monkhorst–Pack *k*-point mesh was used for reciprocal space integrations. Correlation effects were treated by including a Hubbard correlation with $U$ = 2.2 (5.0) eV and $J$ = 0.2 (0.64) eV for Ir (Ti)[32].

**Sample growth.** The superlattice was deposited on a (001)-oriented SrTiO$_3$ single crystal substrate using pulsed laser deposition, with a KrF (248 nm) excimer laser. Before deposition, the substrate was pretreated to have TiO$_2$ termination. The substrate temperature and oxygen pressure were optimized as 700 °C and 0.1 mbar, respectively. Equipped with a reflection high-energy electron diffraction unit, the growth process was in-situ monitored to control the stacking sequence with atomic precision. The details of the growth and sample characterization can be found in Ref. 17.

**Experimental investigation of octahedral rotation pattern.** The enlargement of crystal lattice due to ordered octahedral rotations can be followed by the emergence of half-order Bragg peaks in an x-ray diffraction (XRD) pattern, and the peak intensities are proportional to the octahedral rotation amplitude squared[33,34]. Therefore, from the intensity of specific superlattice peaks, one can identify the lattice distortion due to octahedral rotation. The correspondence of superlattice peak and octahedral rotation has been thoroughly discussed in Refs. [33,34]. To access the weak superstructure due to octahedral rotation, we studied the lattice structure with the synchrotron x-ray source at the 33BM beamline at the Advanced Photon Source of Argonne National Laboratory. A unit cell of $a \times a \times 3c$ ($a$ and $c$ are the pseudocubic in-plane and out-of-plane lattice parameters, respectively) was used to define the reciprocal space notation.

**Magnetization measurements.** Temperature dependent in-plane and out-of-plane remnant magnetizations were measured with a quantum-design superconducting quantum interference device (SQUID) magnetometer.

**X-ray absorption (XAS) and magnetic circular dichroism (XMCD) measurements.** To check the robustness of the $J_{\text{eff}} = 1/2$ model in the present superlattice, we performed XAS and XMCD measurements around the Ir $L_3$ and $L_2$-edges at beamline 4ID-D of the Advanced Photon Source, Argonne National Laboratory.

**Magnetic scattering study.** Resonant magnetic x-ray scattering measurements were performed around the Ir $L_3$-edge at 6IDB at the Advanced Photon Source of Argonne National Laboratory. During the scattering process, a linearly polarized x-ray beam was scattered by both charge and magnetic moments. While the former arises due to non-resonant Thomson scattering, the magnetic scattering intensity can be amplified by choosing an x-ray energy which resonates with the active orbital through atomic transitions. In addition, a charge scattered x-ray has the same polarization as the incident x-ray ($\sigma$-$\sigma$ channel), while the polarization will be rotated 90° by magnetic scattering ($\sigma$-$\pi$ channel). To separate the magnetic contribution from the charge contribution, a polarization analyzer was utilized. This method enables the direct detection of the antiferromagnetic order in a thin epitaxial superlattice less than 50 nm, which would be impossible by other techniques.

**Monte Carlo simulation and fitting parameters.** As derived in the main text, the crossover temperature is given by

$$T_0 = T_{\text{BKT}} + \frac{4b^2 T_{\text{BKT}}}{\left[\ln\left(\frac{c\tilde{J}S^2}{2|J_\perp|S^2 + S|h\,\sin\varphi\,|}\right)\right]^2} \quad (4)$$

To estimate the parameter $b$, we performed the classical Monte Carlo simulation with up to 6144 × 6144 spins for the model

$$H = \sum_{<i,j>} [\tilde{J}\vec{\tilde{S}}_i \cdot \vec{\tilde{S}}_j - \Gamma_1 \tilde{S}_i^z \tilde{S}_j^z \pm \Gamma_2(\tilde{S}_i^x \tilde{S}_j^x - \tilde{S}_i^y \tilde{S}_j^y)] - h \cos\varphi \sum_i \tilde{S}_i^x + h \sin\varphi \sum_i e^{i\mathbf{Q}\cdot\mathbf{r}_i} \tilde{S}_i^y,$$

where +(−) is taken for bonds along the $x(y)$-axis, and $\mathbf{Q} = (\pi, \pi)$ is the antiferromagnetic ordering wave vector, and $\Gamma_1$ and $\Gamma_2$ terms are high-order corrections introduced by the Hund's coupling[18]. In the superlattice, the in-plane rotation angle of octahedra is estimated to be approximately 8°[25], which is a bit smaller than that of bulk $Sr_2IrO_4$. Accordingly, we set $\tan 2\varphi = D/J = 0.27$. For other parameters, we set $\Gamma_1 = 10^{-4} J$ corresponding to out-of-plane gap $\Delta \sim 1$ meV[35,36], and $\Gamma_2 = 5\Gamma_1$[18]. Because the $\Gamma_2$ term does not affect the energy of the canted-spin ground state in the classical spin limit, it is not expected to affect $T_0$. We confirmed that the dependence of the crossover temperature on $\Gamma_2$ is of order $\Gamma_2/\tilde{J}$ and negligible, therefore the $\Gamma_2$ term was omitted in the discussion of the main text. The magnetic field was set down to $h = 2 \times 10^{-5} J$ corresponding to magnetic field ~0.01 T as applied in our experiments. The crossover temperature was calculated in the same way with the experimental results as the inflection point of the antiferromagnetic order parameter as a function of temperature, $\sqrt{\langle |\widetilde{M}_{st}|^2 \rangle}$, where $\widetilde{M}_{st} = \sum_i e^{i\mathbf{Q}\cdot\mathbf{r}_i} \tilde{S}_i^y$.

Above the Berezinskii–Kosterlitz-Thouless transition temperature ($T_{BKT}$), the correlation length diverges as $\xi \sim e^{\frac{b}{\sqrt{t}}}$[37], where $t = \frac{T}{T_{BKT}} - 1$. In the SU(2) symmetric case, it diverges in the vicinity of zero temperature as $\xi \sim e^{\frac{2\pi\rho_s}{T}}$[13], where $\rho_s$ is the spin stiffness. These two expressions should be smoothly connected in the isotropic limit ($T_{BKT} \to 0$), which implies $b \to \infty$. Therefore, the closer to the SU(2) symmetry the interaction is, the faster the divergence of the correlation length becomes. These expressions of the correlation length are valid for both the quantum and the classical spins. Because the quantum nature is expected to be irrelevant with regard to ordering, the difference in the scaling of the correlation length between the quantum and the classical spins is only the renormalization of the energy, namely, the transition temperature or the spin stiffness. We thus expect the parameter $b$ to be independent of the length of spins.

Fitting the numerically obtained data to equation (4), we estimated $T_{BKT}^{cl} \approx 210$ K and $b \approx 3.1$. We used $\tilde{J} \approx 500$ K, which is a common energy scale for magnetic iridates, especially layered systems[28]. Then we fitted the experimental data fixing the value of $b$ and estimated $T_{BKT}^{q} \approx 18$ K and $J_\perp \approx 0.007$ K. In terms of the parameter $c$, which is proportional to the effective energy of vortices, namely the spin stiffness ($c \propto \rho_s$), we obtained $c^{\text{cl}} \sim 1$ for the classical spins and $c^{\text{q}} \sim 10^{-1}$ for the experimental data. Because $\rho_s \propto T_{\text{BKT}}$ at the Berezinskii–Kosterlitz-Thouless transition[38], our estimations are consistent: $c^{\text{q}}/c^{\text{cl}} \approx T_{\text{BKT}}^{\text{q}}/T_{\text{BKT}}^{\text{cl}} \sim 10^{-1}$. The reduction of the transition temperature can be explained by the quantum fluctuations and possible disorder in the real material.

**Acknowledgments** The authors acknowledge experimental assistance from H. D. Zhou, E. Karapetrova, C. Rouleau, Z. Gai, J. K. Keum, and N. Traynor. The authors would like to thank E. Dagotto, I. Zalzinyak, D. McMorrow, J.-H. Chu and H. D. Zhou for fruitful discussions. J.L. acknowledges the support by the start-up fund and the Transdisciplinary Academy Program at the University of Tennessee. J.L. and H.X. acknowledge the support by the Organized Research Unit Program at the University of Tennessee and the support by the DOD-DARPA under Grant No. HR0011-16-1-0005. M.P.M.D. and D.M. are supported by the U.S. Department of Energy, Office of Basic Energy Sciences, Early Career Award Program under Award Number 1047478. H.S. and C.D.B. are supported by funding from the Lincoln Chair of Excellence in Physics. D.K. and L.H. acknowledge the support by the ERDF (project CZ.02.1.01/0.0/0.0/15_003/0000485) and the Grant Agency of the Czech Republic Grant (14-37427G). A portion of the work was conducted at the Center for Nanophase Materials Sciences, which is a DOE Office of Science User Facility. Use of the Advanced Photon Source, an Office of Science User Facility operated for the U. S. DOE, OS by Argonne National Laboratory, was supported by the U. S. DOE under Contract No. DE-AC02-06CH11357.

**Author contributions** C.D.B., M.P.M.D. and J.L. conceived and directed the study. L.H., D.M., J.Y. and C.F. undertook sample growth and characterization. L.H., D.M., J.Y., J.W.K, and P.J.R. performed magnetic scattering measurements. L.H., D.M., G.F., Y.S.C., and D.H. conducted XMCD measurements. L.H., D.M., J.Y., L.H. and D.K. collected synchrotron XRD data. L.H. and J.L. analyzed data. H.S. and C.D.B. performed Monto Carlo simulations. T.R.D. and H.X. performed first principles calculations. L.H., H.S., C.D.B., M.P.M.D. and J.L. wrote the manuscript.

**Additional information** Supplementary information is available in the online version of the paper. Reprints and permissions information is available online at www.nature.com/reprints. Publisher's note: Springer Nature remains neutral with regard to jurisdictional claims in published maps and institutional affiliations. Correspondence and requests for materials should be addressed to H.X. (hxu8@utk.edu), C.D.B. (cbatist2@utk.edu), M.P.M.D. (mdean@bnl.gov) or J.L. (jianliu@utk.edu).